\documentclass[journal=ancac3,manuscript=article,layout=traditional]{achemso}

\usepackage{graphicx}
\usepackage{amsmath}
\usepackage[version=3]{mhchem}
\usepackage[usenames]{color}
\usepackage{physics}
\usepackage{xr} 
\usepackage[colorlinks=true, linkcolor=black, citecolor=black, urlcolor=black, filecolor=black,pdftex]{hyperref}
\usepackage{orcidlink}

\newcommand{\figref}[1]{Fig.~\ref{#1}}
\newcommand{\sfigref}[1]{Fig.~\ref{#1}}

\newcommand{\PIR}{D$_{0}^{+}$}
\newcommand{\GS}{S$_{0}$}
\newcommand{\ESa}{S$_{1}$}
\newcommand{\ESx}{S$_{1,x}$}
\newcommand{\ESy}{S$_{1,y}$}
\newcommand{\ESb}{S$_{2}$}
\newcommand{\TBP}{H$_{2}$TBP}
\newcommand{\Pc}{H$_{2}$Pc}

\newcommand{\To}{TBP$_{111}$}
\newcommand{\Tz}{TBP$_{110}$}
\newcommand{\Po}{Pc$_{111}$}

\usepackage[normalem]{ulem}

\title{Tunable Luminescence From a Single Free-Base Porphyrin Molecule By Controlled Access to Optically Active States}

\author{Eve Ammerman\,\orcidlink{0000-0003-3588-2245}}
\email{eve.ammerman@empa.ch}
\affiliation{nanotech@surfaces Laboratory, Empa -- Swiss Federal Laboratories for Materials Science and Technology, D\"ubendorf 8600, Switzerland}

\author{Nils Krane\,\orcidlink{0000-0002-8273-6390}}
\affiliation{nanotech@surfaces Laboratory, Empa -- Swiss Federal Laboratories for Materials Science and Technology, D\"ubendorf 8600, Switzerland}

\author{Bruno Schuler\,\orcidlink{0000-0002-9641-0340}}
\affiliation{nanotech@surfaces Laboratory, Empa -- Swiss Federal Laboratories for Materials Science and Technology, D\"ubendorf 8600, Switzerland}

\begin{document}
\thispagestyle{empty}

\begin{center}
    \date{\today}
\end{center}

\begin{abstract}
Scanning tunneling microscopy-induced luminescence (STML) provides access to optical properties of individual molecules through a cascade of relaxation processes between many-body states. Insufficient charge attachment energies quench the relaxation cascade via optically excited states, causing even intrinsically bright molecules to remain dark in STML. Here, we utilize substrate work function control and tip-induced gating of the double barrier tunnel junction to induce an energy shift of the ionic transition state of a single free-base tetrabenzoporphyrin (\TBP) to control access to optically excited states and bright exciton emission. The experimental observations are validated by a rate equation and polaron model considering the relaxation energy of the NaCl decoupling layer upon charging of the molecule.
\end{abstract}

\maketitle
\setcounter{page}{1}

\textbf{Keywords:} Electroluminescence, Scanning Tunneling Microscopy, Charge Dynamics, Electronic Transitions, Tunnel Junctions, Molecules, Porphyrin

\newpage

\section{Introduction}
Understanding the optoelectronic (photophysical) properties of organic molecules, such as porphyrins, is technologically relevant for organic light emitting diodes (OLEDs) and organic photovoltaics~\cite{Mathew2014}, and fundamentally important for understanding biological systems, such as photosystem I \& II~\cite{Jordan2001,Umena2011}. Recent developments in on-surface synthesis methods have led to the incorporation of porphyrins alongside $\pi$-radical magnetic nanographenes~\cite{Mateo2020,Sun2021,Chen2024,xiang2025zigzag}, resulting in tailor-made molecular structures with engineered electronic, magnetic, and optical properties. Porphyrins offer unique opportunities for optical and spin tunability through metalation of the core and coordination chemistry, thereby enabling exciting prospects for designing the coupling in hybrid porphyrin-nanographene structures.

Scanning tunneling microscopy (STM) provides access to single molecules and orbital resolved electronic measurements. Additionally, STM luminescence (STML) allows probing of their optical properties with characterization comparable to both fluorescence and Raman spectroscopy~\cite{kuhnke2017atomic,doppagne2018electrofluorochromism,roslawska2022mapping,jiang2023topologically}. Phthalocyanine (\Pc) chromophores, structurally analogous to free-base tetrabenzoporphyrins (\TBP), have become the benchmark system in STML studies~\cite{zhang2016visualizing,imada2016real,doppagne2017vibronic,zhang2017sub,doppagne2018electrofluorochromism,cao2021energy,roslawska2022mapping,dolezal2024single}. In contrast, nanoscale optical characterization of porphyrins remains surprisingly scarce~\cite{deng2002stm,liu2009stm,zhu2013self,dong2010TPP,chong2016}, despite their central role in surface science~\cite{gottfried2015surface}.

Although \TBP\ and \Pc\ share a benzene‑annulated tetrapyrrolic framework, differences in the macrocycle composition lead to distinct photophysics~\cite{lu2016optically}. In \Pc, the meso positions are occupied by nitrogen atoms instead of carbon, increasing electronic delocalization and rigidity~\cite{stuzhin1996}. Consequently, they exhibit different excited‑state relaxation dynamics, although both families generally maintain high fluorescence quantum yields~\cite{lu2016optically}.

Here we show that \TBP\ are inherently dark in single-molecule electroluminescence on 3ML NaCl/Ag(111) due to unfavorable energy alignment. \TBP\ has a smaller hole attachment energy than the chemical analog \Pc, suppressing STML emission by $\sim$98\,\% compared to \Pc\ on Ag(111). By locally tuning the electrostatic environment, either via substrate work function or tip-induced gating in a double-barrier tunnel junction, we can overcome this limitation; accessing otherwise forbidden optical transitions and activating bright emission from individual \TBP\ molecules.

\section{Results and Discussion}
\subsection{Molecular ionic transition state energy on few layer NaCl}
\begin{figure}[h]
\includegraphics[width=\textwidth]{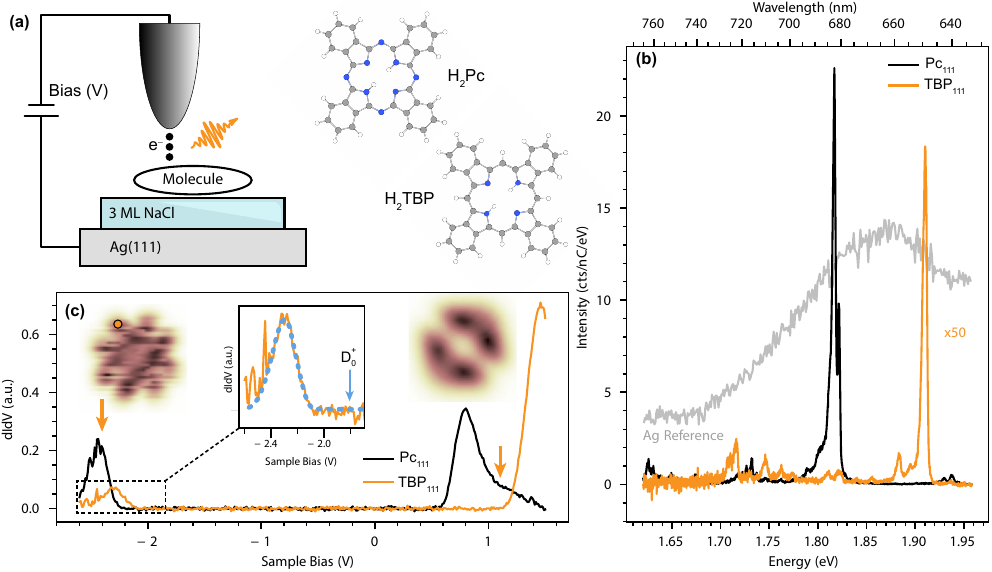}
\caption{\label{1}\textbf{Electronic and optical fingerprints of \TBP} (a) Cartoon of STM luminescence experiments of \Po\ and \To\ adsorbed on a thin-film NaCl decoupling layer and supported by Ag(111). (b) Comparison of fluorescence intensity from single \To\ and \Po\ molecules. STML of \To\ measured at -2.3\,V and \Po\ measured at -2.5\,V. (c) Scanning tunneling spectroscopy of \TBP\ shown with comparison to \Po\ to emphasize the upward shift of the molecular transport resonances. Constant-height STM images show the spatial distribution of the PIR and NIR recorded at -2.4\,V and 1.1\,V, respectively. In both STS measurements, feedback was opened at positive sample bias with a setpoint current of 40\,pA. The STS and STML data in panel b and c were recorded from molecules adsorbed on the same NaCl island with the same tip to ensure fair comparison of optical intensity. The inset in panel c shows a polaron model fit to the PIR, yielding a \PIR\ state energy of 1.81\,eV.}
\end{figure}

STML measurements are performed on individual \TBP\ and \Pc\ molecules decoupled from a Ag(111) substrate by 3 ML insulating NaCl islands, as sketched in \figref{1}a. We will refer to these two systems as \To\ and \Po\ in the following. \figref{1}b shows the corresponding STML spectrum of \Po\ (black) recorded with an excitation rate of 63\,pA at -2.5\,V and acquisition time of 30\,s. The dominant peak at 1.81\,eV and its lower energy vibronic satellites can be assigned to the \ESa\ (or $Q_x$) emission line and the weaker spectral lines around 1.94\,eV to the \ESb\ (or $Q_y$) emission, in agreement with literature~\cite{imada2016real}. The STML spectrum on the nearby \TBP\ molecule (orange), taken with the same tip and on the same NaCl island, shows a peak at 1.91\,eV, which we assign to \ESa, and no emission at higher energy could be detected. Strikingly, the emission intensity of \To\ is significantly weaker than the \Po\ emission, necessitating a higher excitation current (150\,pA) and relatively long integration time (120\,s). After normalizing the luminescence spectra to photon counts per tunneled electron, the \To\ spectrum has 50 times less intensity than the \Po\ spectrum. Such a low photon yield of \To\ poses a significant challenge for experiments requiring high sensitivity, such as high-resolution fluorescence mapping~\cite{roslawska2022mapping}, detection of weak vibronic satellite emission~\cite{doppagne2017vibronic,Dolezal2022,vasilev2024}, or potential phosphorescence on metalated porphyrins~\cite{grewal2025}.

To explain the STML excitation mechanism and dark emission of \To\ we adopt the many-body description of molecular electronic states which was first introduced by Miwa et al. and is now a well established practice which provides a comprehensive description of electronic excitations and optical emission arising from different charge states.~\cite{Miwa2019,Hung2023bipolar,Jiang2023,kaiser2025}. In the many-body framework, the energy level alignment between a transient charge state and the optically excited state should be considered to be the primary factor restricting fluorescence of \To. Scanning tunneling spectroscopy (STS) measurements of \To\ and \Po\ (\figref{1}c) were recorded at the same tip positions as the STML spectra in \figref{1}b. The STS spectra reveal indeed a pronounced upward shift of the positive and negative ion resonances of \To\ relative to \Po, accompanied by an increased transport gap. This shift can be attributed to the presence of four additional electronegative nitrogen atoms in the \Pc\ macrocycle.

At first glance, the observed shift does not appear to limit electro-excitation of \To. The onset of the positive ion resonance (PIR) in the STS spectra appears at an applied bias of –2.1\,V suggesting a nominal energy of 2.1\,eV of the positively charged molecule (\PIR) upon hole injection. This energy lies approximately 200\,meV above the optical emission line at $\sim$1.91\,eV, seemingly satisfying the energy requirement for electroluminescence. However, this interpretation neglects the relaxation processes accompanying molecular charging on the surface. In particular, the ionic NaCl surface has been shown to substantially contribute to the reorganization energy of the system during charge injection into the molecule~\cite{Repp2005,Fatayer2018}. This high electron-phonon coupling of the NaCl environment leads to vibronic excitation and, consequentially, to an apparent broadening and shift in energy of the the ion resonances in STS. The amount of the energy shift is comparable to the charge-state relaxation energy, $\lambda_{+}$, as illustrated in \sfigref{fig:polaron}. To obtain the true zero-phonon energy of \PIR, the precise onset of the ion resonance peak needs to be determined, which is experimentally challenging. At such high electron-phonon coupling, the zero-phonon transition becomes vanishingly small, due to an effective Franck-Condon blockade~\cite{kochFranckCondon2005} and increasing the tunneling current signal by approaching the tip risks a picking-up of the molecule. Vasilev at al. demostrated an alternative method to determine the true ground-state energy of a charged molecule on NaCl by utilizing polaron theory to model the lineshape of the broadened resonance~\cite{vasilev2024}. We fitted the experimental STS spectra with the polaron model adapted from Ref.~\citenum{vasilev2024} and obtained a \PIR\ energy of 1.81\,eV for \To, after accounting for a voltage drop of 10\,\%. For \Po, the \PIR\ energy was found to be of 1.98\,eV (more details in the SI). The results suggest the \PIR\ energy of \Po\ is about 200\,meV higher than its \ESa\ state at 1.81\,eV, enabling the \PIR\,$\rightarrow$\,\ESa\ transition and, accordingly, bright emission from \Po. On the other hand, for \TBP, the \PIR\ energy is $\sim$100\,meV lower than the respective \ESa\ state and emission should be fully quenched. To explain the experimentally observed STML intensity, one needs to consider the dielectric and geometric properties of the STML tunnel junction.

\subsection{Tuning the local electrochemical potential}
In a double-barrier tunneling junction (DBTJ) geometry, a fraction of the applied bias voltage drops across the insulating film, leading to a shift in the electrochemical potential ($\Delta E$) of the adsorbed molecule with respect to the substrate~\cite{NazinPNAS2005}. For molecules on few-layer (3-4 ML) NaCl, a voltage drop of $\alpha\sim$10\% can be estimated~\cite{Imada2018voltagedrop,Miwa2019,vasilev2024}. The STML measurements are typically performed with applied absolute voltages in the range of 2-3\,V providing an effective gating voltage of 200-300\,meV, which is on the order of the energy difference between the \To\ \PIR\ and \ESa\ states. Hence, this gating effect is expected to lift the \PIR\ state energy across the transition threshold to \ESa\ leading to an emission intensity strongly dependent on the applied bias.

\begin{figure}[h]
\includegraphics[width=\textwidth]{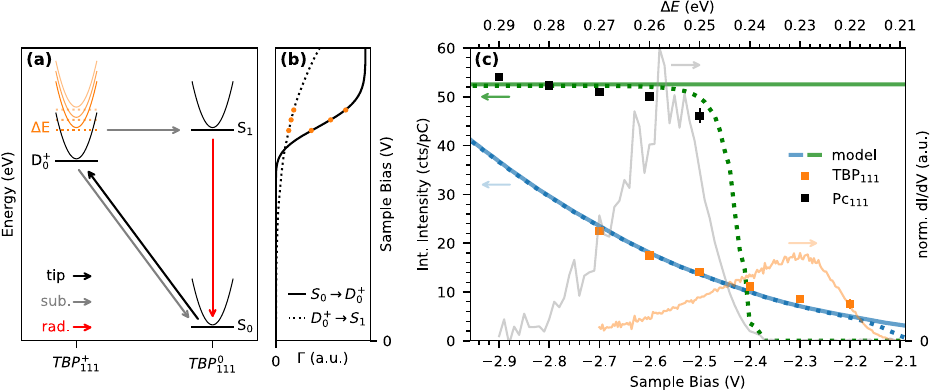}
\caption{\label{2}\textbf{Local tip gating in single molecule fluorescence} (a) Many-body diagram describing the electronic states and transitions of a \To\ molecule during STML experiments. Black and grey arrows connecting different \To\ charge states represent electron transfer involving the tip and substrate while the red arrow represents radiative relaxation. The excess relaxation energy needed for charge transport is depicted as a parabola above each state. A voltage dependent shift in \PIR\ energy, $\Delta E(V)$, is present during experiments and indicated by a series of orange dashed lines. (b) Voltage dependent \GS\ $\rightarrow$ \PIR\ and \PIR\ $\rightarrow$ \ESa\ transition rates. Orange circles correlate with $\Delta E(V)$ shown in (a). (c) [left axis] Integrated intensity of the \ESa\ spectral feature for both \To\ and \Po. Spectral intensity is normalized by the tunneling current and integration time. The raw spectra and integration ranges are shown in supplementary \sfigref{fig:raw_stml}. A model simulation of the voltage dependence is shown for \To\ in blue and \Po\ in green. A dashed colored line represents the model simulation with a dark background current of 150\,fA that does not contribute to the emission but reduces the observed emission rate at low currents. [right axis] Normalized dI/dV of \To\ and \Po\ PIR shown for reference. For the d$I$/d$V$ measurements, the STM feedback was open with a setpoint sample bias of -2.7\,V and current -50\,pA for \To\ and -2.9\,V and -30\,pA for \Po.
}
\end{figure}

Experimentally, the effect can be observed by taking STML spectra at different bias voltages but constant tip height, to keep the tip's nanocavity plasmon unchanged~\cite{Hung2023bipolar,kaiser2025}.
With increasing absolute bias voltage, more vibronic tunneling channels from \GS\ to \PIR\ are continuously opened, leading to an increase in current and, accordingly, STML intensity~\cite{Miwa2019}. To account for this effect, the STML intensity was normalized by the total current.
The normalized photon counts per tunneled electron emission intensities of the \ESa\ line were integrated between 642\,nm - 652\,nm and are displayed in \figref{2}c. Instead of a sharp step like increase, once the gating voltage is lifting \PIR\ energetically above \ESa, there is a steady increase in intensity with applied bias voltage. This dependence can be understood when we examine the relative transition rates between the \PIR\ charge state and the optical state \ESa. \figref{2}a illustrates the many-body state diagram relative to the ground state (\GS). As described above, the \PIR\ state energy on Ag(111) lies about 100\,meV below the \ESa\ state without any bias applied, preventing the transition and subsequent radiative relaxation. With a negative bias voltage applied to the STM junction \PIR\ is shifted upwards by $\Delta E(V)$ due to the voltage drop, enabling access to the \PIR\ $\rightarrow$ \ESa\ transition. Analogous to the \GS\ $\rightarrow$ \PIR\ transition, \PIR\ $\rightarrow$ \ESa\ involves a charge change of the molecule that is subject to the strong electron-phonon coupling of the NaCl layer. The transition step is accordingly shifted to higher energies and broadened, leading to the steady increase in intensity with bias voltage, as the continuous shift of \PIR\ steadily increases the transition rate into \ESa.

Using a simple rate equation model~\cite{kaiser2025}, we simulated the voltage dependent emission rate to reproduce the experimental observation and strengthen the gating model.
For this, the voltage-dependent transition rates $\Gamma_{S_0 \rightarrow D^+_0}$ and $\Gamma_{D^+_0 \rightarrow S_1}$ were calculated with the polaron model described above. The critical role of the polaron model is to capture the vibronic response of the NaCl upon (dis)charging of the molecule and provide a voltage dependent spectral function $S(V) \propto \mathrm{d}\Gamma/\mathrm{d}V$ that can be used to describe the transition rates $\Gamma(V)$. The STS measurement of the positive ion resonance largely reflects the voltage dependence of the transition rate derivative between \GS\ and \PIR, thus, $\Gamma_{S_0 \rightarrow D^+_0}$ can be well approximated by the integral of the polaron spectral function fitted to the experimental $\mathrm{d}I/\mathrm{d}V$. However, there is no direct measure of the gating dependent transition between the charge state and \ESa. As an approximation, we assumed the relaxation energy $\lambda_+$ for both transitions to be the same and described them with the same spectral function, depending on the effective applied voltage between molecule and corresponding lead, i.e. $S((1-\alpha)V_b)$ and $S(-\alpha V_b)$, respectively. The resulting voltage dependent transition rates are sketched in Fig.~\ref{2}b and \sfigref{fig:VD_transition_rates}. Because both the tip height and lateral position strongly influence the local voltage drop and tunneling rate, these two quantities were treated as fitting parameters in the model. The resulting shift in the electrochemical potential $\Delta E$ (top axis of \figref{2}c) scales with a voltage drop of 10.2\,\% which is within the expected range of 10-15\,\%. More details on the model and its implementation are provided in the SI including a detailed analysis regarding the voltage drop dependence of the model. The resulting simulated voltage dependence of the emission rate (blue line in \figref{2}c) is in good agreement with the experiment (orange dots).

To validate the gating model, we also recorded the voltage dependent emission of \Po\ (black dots in Fig.~\ref{2}c). Contrary to \To, the normalized intensity remains near constant with the bias voltage. Due to the higher ion resonance energy of \Po, the \PIR\ $\rightarrow$ \ESa\ transition is almost "fully opened" and the STML intensity is only limited by the initial \GS\ $\rightarrow$ \PIR\ transition. In the typical STM regime, the latter is proportional to the tunneling current and normalizing the STML intensity with current will yield a near constant signal, as observed in experiment. Please note, for very small currents, i.e. at bias voltages  $>$-2.6\,V where tunneling into the ion resonance is limited, one needs to consider the effect of a small background current $I_\mathrm{bg}$, flowing directly from the sample towards the tip. It skews the normalization of the STML intensity, as the additional background current becomes significant for very small total currents in the denominator of the normalization $J_\mathrm{ph}/(I_\mathrm{mol}+I_\mathrm{bg})$ and causes an apparent downshift of the STML intensity. To capture this effect in our model, we added a constant offset of 150\,fA to the simulated current $I_\mathrm{mol}$ (dotted lines in Figure~\ref{2}c).

In the following, we discuss the applicability of other excitation mechanisms, reported in literature. Direct plasmonic excitation has been reported for molecules on thin decoupling layers, where electrons tunneling directly between tip and substrate can transfer energy to the molecule, mediated by the plasmonic states in the nanocavity~\cite{doppagne2017vibronic}. The process requires sufficiently high currents between tip and substrate and has, for this reason, been shown for bias voltages within the electronic gap of the molecule. For \To, the effect can be ruled out for two reasons: on the one hand, there is little to no current flowing directly between tip and substrate, as can be seen by the vanishing current before the onset of the \PIR. On the other hand, plasmonic excitation should occur already at absolute bias voltages larger than the photon energy of 1.91\,V, which has not been observed in experiment.
Another excitation mechanism could be electronic pumping via the triplet state $T_1$ into, either, a higher positive ion resonance or a negatively charged state~\cite{Luo2024upconversion,kaiser2025}. These processes require tunneling of two electrons between tip and molecule to reach \ESa\ and are expected to be much smaller than the single-electron gating model explained above. Electronic pumping alone could not be fitted with the rate equation model to the experimental data within a physically reasonable parameter space.

To recapitulate, during our STML measurements two electronic excitations are required in order to access the optically active states, \GS\ $\rightarrow$ \PIR\ and \PIR\ $\rightarrow$ \ESa. Each excitation can be treated as independent charge state transition and both are subject to a resonance shift and broadening due to the underlying NaCl layer. Therefore, despite a shift of $\Delta E > 200$\,meV for bias voltages at the onset of the PIR, luminescence remains below the experimental detection limit. It is not sufficient to simply raise the \PIR\ energy above \ESa\ but a larger shift is needed to compensate for said resonance shift and broadening. Extrapolation of the rate equation model to more negative bias voltages shows, that the emission intensity of \To\ on Ag(111) is still quenched by 94\,\% for bias voltages around -2.5\,V. The model indicates a bias voltage $<$ -5.6\,V, $\Delta E \ge  560$\,meV, is needed to achieve optimal energy level alignment. Applying such a large bias voltage in an experiment would lead to unwanted effects, including a transition from the tunneling regime to field emission which would result in a loss of spatial resolution and risk damage to the molecule being studied.

\subsection{Engineered energy level alignment for optical characterization}
\begin{figure}[h]
\includegraphics[width=\textwidth]{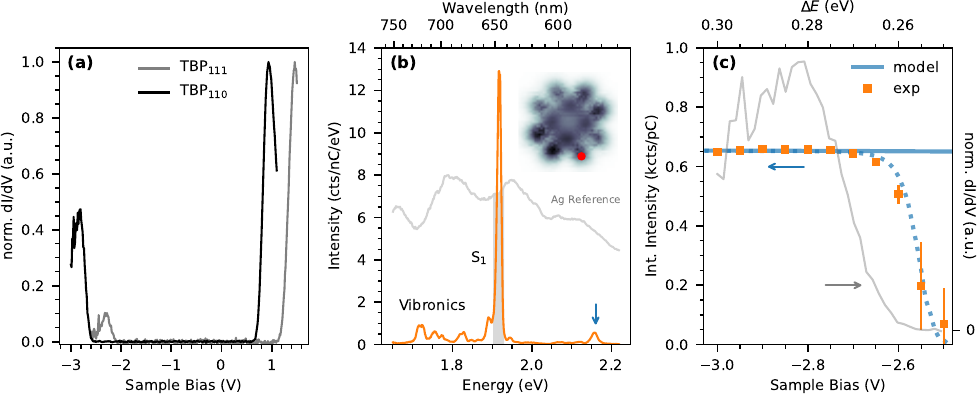}
\caption{\label{3}\textbf{Engineered energy level alignment for enhanced optical characterization} (a) Comparison of STS measurements for TBP on NaCl with Ag(111) and Ag(110) as the conducting substrates. The STS measurement with Ag(110) was recorded with the feedback opened with a sample bias of -3\,V and current of -55\,pA. (b) STML spectra of a \Tz\ molecule. Measurement performed with $V$ = -3.0\,V, $I$ = -314\,pA and $t_\text{acq}$ = 15\,s. A reference spectrum recorded on the Ag surface at $V$ = -3.0\,V is shown in grey. An inset with a constant-height STM image indicates where the spectrum was recorded. A blue arrow indicates the observation of a new spectral feature at 2.16\,eV. (c) [left axis] Integrated intensity of the \ESa\ spectral feature indicated in (b). Spectral intensity is normalized by the tunneling current and integration time. The raw spectra and integration ranges are shown in supplementary \sfigref{fig:raw_stml}. A model simulation of the voltage dependence is shown for \Tz\ in blue. A dashed colored line represents the model simulation with a dark background current of 150\,fA. [right axis] Normalized dI/dV of the \Tz\ PIR shown for reference. Error bars are provided to show the rapid decrease in signal to noise when transport through the PIR is minimal.}
\end{figure}

For reasons previously mentioned, the chemical potential shift induced by the voltage drop in the STM junction provides only limited tuning range. The local gating effect is insufficient to fully overcome the energetic constraint limiting the \PIR\ $\rightarrow$ \ESa\ transition, as shown by the strong voltage dependence of the emission intensity. The effective voltage drop could be increased by introducing a thicker decoupling layer. However, the increased lever arm of the gating potential would be even more sensitive to the tip position relative to the molecule, making the understanding of experiments that include spatially mapping the STML intensity more complex. This effect can already been seen for 3\,ML as discussed below. To extend the tuning range of the chemical potential, we exploit the intrinsic energy level alignment defined by the substrate work function ($\Phi$), which enables a static shift of the molecular ion resonances. To this end, we performed STML measurements of \TBP\ on NaCl films deposited on Ag(110) with a thickness of $\sim598$\,pm, which will be referred to as \Tz. The decoupling layer thickness was chosen to achieve similar charge state lifetimes, and ensure comparable decoupling, between \Tz\ and \To\ (see Figure~\ref{fig:approach_curves}). The Ag(110) substrate retains the strong plasmonic enhancement of silver while reducing the work function from $\Phi_\mathrm{Ag111}=4.5$\,eV to $\Phi_\mathrm{Ag110}=4.1\,$eV~\cite{Derry2015}. Lowering the workfunction by 400\,mV is expected to shift the positive ion resonance energies of adsorbed molecules by a similar amount to higher energies. This shift is confirmed experimentally by comparing the STS spectra of decoupled \TBP\ on Ag(111) and Ag(110), with the PIR on Ag(110) being shifted by 400\,meV to -2.5\,eV (\figref{3}a). A polaron model fit of the STS spectra of \Tz\ (\sfigref{fig:polaron}b) yields a \PIR\ energy of 2.25\,eV, matching the work function shift. The shift in the \PIR\ state energy provides approximately $\sim$300\,meV excess energy relative to the \ESa\ state, comparable to the \PIR\ $\rightarrow$ \ESa\ transition of \Po. Correspondingly, STML measurements of \Tz\ reveal a significant increase in the emission intensity and the appearance of an additional emission line at
2.15\,eV, or $230$\,meV above \ESa\ (\figref{3}b). The new emission line can be tentatively assigned to \ESb, which has been reported to be $\sim150$\,meV above \ESa\ for absorption spectroscopy in vitrified solution~\cite{Kuzmitskii1984}. Another origin of the emission line could be hot-luminescence, where a vibrationally excited molecule emits while relaxing into the vibrational ground state~\cite{chong2016,Rai2023}.

A direct comparison of emission intensity with \Pc\ is not possible on the NaCl/Ag(110) substrate, since the reduced workfunction causes a charging of \Pc\ and thus prohibits a reliable comparison. To corroborate the higher STML efficiency of \Tz\ compared to \To\ we again performed voltage dependent STML spectroscopy. In \figref{3}c, the voltage dependence of the normalized emission intensity is shown. It exhibits a clear plateau as soon as the applied voltage reaches the onset of the PIR, in contrast with the observations for \To, where the emission increases gradually, even at voltages significantly beyond the PIR (\figref{2}c). By engineering the level alignment we shift the \PIR\ sufficiently high in energy that the \PIR\ $\rightarrow$ \ESa\ transition becomes "saturated" and does not increase anymore with more negative bias voltages (\sfigref{fig:MBD_TBP110}). Applying the rate-equation model to the STML data reproduces an approximately constant emission rate for \ESa, similar to the one observed for \Po.

\subsection{Gating effects in spatially resolved fluorescence maps}
A significant benefit of performing STML measurements on single molecules is the angstrom-scale spatial resolution provided by the technique. In hyper-resolved fluorescence mapping (HRFM), many fluorescence spectra are recorded on a dense grid across an individual molecule. Analysis of the intra-molecular spatial distribution of spectral features can help identify the origin of emission lines and and provide insight into the photophysics and vibronic fine structure of molecules with distinct structure and adsorption geometries~\cite{doppagne2017vibronic,vasilev2022,doppagne2020tautomerization}.

Time-dependent density functional theory (TD-DFT) calculations of the \TBP\ excited states predict an \ESa\ transition dipole along the axis $A_x$, defined by the two inner hydrogens of the macrocycle (see \sfigref{fig:simMaps_TD}), while the energetically higher state, \ESb\, has a transition dipole along the orthogonal $A_y$ axis, analogous to \Pc. In experiment, we expect a mixing of the axes due to fast tautomerization of the two central hydrogen atoms. Previous STML studies of \Pc\ report an NaCl substrate-induced anisotropy can lift the degeneracy between the corresponding excited states \ESx\ and \ESy~\cite{doppagne2020tautomerization,roslawska2022mapping}. This splitting is reflected in the double peak structure of the main emission line of \Po\ shown in Fig.~\ref{1}b.

\begin{figure}[h]
\includegraphics[width=\textwidth]{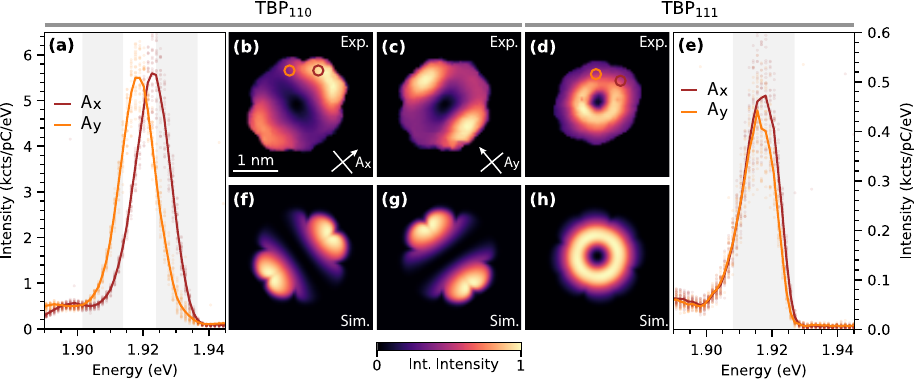}
\caption{\label{4}\textbf{Fluorescence mapping of the \ESa\ emission line of \TBP} (a) STML spectra from the $A_x$ and $A_y$ axes of \Tz, at locations indicated in (b). (b,c) \ESa\ constant height fluorescence maps of the two tautomer states of \Tz\ between 1.923-1.937\,eV and 1.902-1.914\,eV, respectively. The orientation of the optical state matches the $A_x$ and $A_y$ symmetry axes of the molecule as indicated by the white arrows. (d) Constant height fluorescence map of \To\ between 1.907-1.927\,eV. (e) \To\ reference STML spectrum from the $A_x$ and $A_y$ axes. (f,g) Simulated STML maps of \ESa\ of the \Tz\ tautomers. (h) Simulated STML map of the \To\ molecule. STML map simulations were created using the polaron and rate model described in the main text with a spatially dependent voltage drop described in the supplement. All STML map intensities are normalized to 1 and share the same scale bar. The \Tz\ map used for (b) and (c) was recorded at -2.8\,V and $t_\text{acq}$ = 15\,s. The \To\ map used for (d) was recorded at -2.4\,V and $t_\text{acq}$ = 30\,s. The $A_x$ and $A_y$ reference spectrum are averages of 29 measurements that were recorded within a 6 pixel diameter indicated by the colored circles in (b) and (d). The integration regions used to generate fluorescence maps in are indicated with grey boxes in (a) and (e). Post-processing of the experiential STML maps, (b)-(d), with a $3\times3$ median filter is performed to remove the effect of cosmic rays and improve image quality.
}
\end{figure}

A similar splitting of the two states can also be observed for \Tz. Figure~\ref{4}a shows two emission spectra taken along the $A_x$ and $A_y$ axes of \Tz\ exhibiting a relative lineshift of $\sim10$\,meV. 
Utilizing the splitting of the emission lines, we can map the spatial intensity distribution of \ESx\ and \ESy, by integrating the spectral intensities on the high and low energy flanks of the primary emission lines (see grey areas in \figref{4}a). The resulting fluorescence maps, normalized by excitation current, are shown in \figref{4}b,c. They both feature a nodal plane with orthogonal symmetry axes and maximal emission intensity near the periphery of the molecule. These experimental HRFM maps are well reproduced by TD-DFT calculations of the corresponding excited states, including the plasmonic response of the STM tip in the simulation (see \figref{4}f,g and \sfigref{fig:simMaps_maps}a). 

For \To, in contrast, we do not observe a splitting of the two tautomer excited states. The STML spectra shown in \figref{4}e exhibit a slight intensity difference between the $A_x$ and $A_y$ directions, likely reflecting small variations in the tautomer lifetime, but no distinct lineshift.
The HRFM of the integrated primary emission line (grey area in \figref{4}e), is shown in \figref{4}d. Consistent with the near degeneracy of \ESx\ and \ESy, the map does not exhibit the two-fold symmetry that was observed for \Tz, but instead displays a ring-like intensity distribution. The stronger splitting of \ESx\ and \ESy\ for \Tz, compared to \To, can be rationalized by the stronger anisotropy of the NaCl/Ag(110) interface, where significant structural anisotropy of the NaCl layer has been reported~\cite{quertite2017nacl}.

Besides the tautomerization effects discussed above, comparison between the HRFM of \Tz\ and \To\ shows an additional difference in the spatial distribution of emission intensity. For \Tz, the regions of highest normalized intensity are located at the edges of the map, whereas for \To\ the \ESa\ emission forms a ring close to the center of the molecule. Since fluorescence maps primarily reflect the coupling between the plasmonic nano-cavity and the excited-state transition density~\cite{roslawska2022mapping,Ferreira2025}, both systems would \textit{a priori} be expected to exhibit similar spatial profiles. To understand the difference, we need to consider the effect of the voltage drop, and subsequent gating, that significantly influences excitation efficiency in \To. The non-linear behavior shown in \figref{2}c is intrinsically related to the energy shift of the \PIR\ state, which is proportional to the voltage drop, and is highly sensitive to the tip position~\cite{Pradhan2005,Nazin2005spatialcharging}. The tip-induced gating of the molecule is highest when the tip is placed above the center of the molecule and decreases when the tip is moved radially away from the molecule center. In the case of \To, the reduced gating potential when the tip is placed at the edge of the molecule limits the \PIR\ $\rightarrow$ \ESa\ transition rate and significantly reduces the emission intensity. On the other hand, the transition density of \TBP\ does not couple to the tip plasmon when the tip is centered above the molecule. These two effects taken together lead to vanishing intensity in both the center of the molecule as well as at the edges, leading to a ring shaped emission profile observed in the \To\ fluorescence maps.

To corroborate our explanation of the observed differences in the spatial profile of STML emission of \To\ and \Tz\, we adapt our polaron and rate equation model to spatially resolved STML map simulations by expanding the voltage drop to include a tip-position dependency. We assume a simple two-point-charge model to describe the spatial dependence of the gating~\cite{Krane2019} and the coupling to the plasmon~\cite{Ferreira2025} (see SI for details). The simulated fluorescence map for \To, including fast tautmerization effects, is shown in \figref{4}h reproducing the experimentally observed ring-shaped intensity profile. 

\section{Conclusions}
In summary, we have shown that tuning the hole-attachment energy via tip-induced gating or by controlling the substrate work function enables efficient electrical excitation into optically active states from single \TBP\ molecules. Quantitative analysis of the hole attachment energy using a polaron model highlights the critical role of the relaxation energy of the NaCl decoupling layer in determining the energy requirements for transitions into the excited states. We demonstrated the energetically restricted access to the optical excited state \ESa\ for \To\ by voltage-dependent STML emission, which could be modeled by rate equations that consider the NaCl relaxation energy extracted from STS. Comparison of \To\ and \Tz\ fluorescence maps clearly demonstrate the significance of local gating effect and the impact it has on the spatial characterization of electroluminescence from single molecules. 
Controlled energy-level engineering in STML enables a detailed, quantitative probe of molecular photophysics in regimes where the ionic transport energy and optical excitations are comparable, offering a general strategy for studying chromophores at the single-molecule level.

\section{Methods}
\subsection{Sample Preparation}
Sample preparation of \TBP\ and \Pc\ decoupled from Ag(111) and Ag(110) by thin-films of NaCl. Sputter and anneal of single-crystal silver substrates by repeated argon ion bombardment and heating to 673\,K. Growth of NaCl thin-films on Ag surfaces at room temperature by sublimation from a Knudson-type effusion cell at 740\,°C and post-annealing to 200\,°C for NaCl on Ag(111) and 230\,°C for NaCl on Ag(110). The difference in NaCl thickness is chosen to achieve comparable topographic heights in STM measurements of $\sim$600\,pm and comparable charge state lifetimes (see Figure~\ref{fig:approach_curves}). In-situ flash-deposition of molecules was performed from a direct-current Si heater with direct line of sight to the STM scan head held at 4.5\,K.

\subsection{STML Measurements of Single Molecules}
SPM measurements were performed using a CreaTec Fischer \& Co. GmbH scanning probe microscope at liquid helium temperatures (\(T < 5 \, \text{K}\)) under ultrahigh vacuum (\(p < 2 \times 10^{-10} \, \text{mbar}\)). STM and STS measurements performed with a FEMTO-DLPCA200 preamplifier with a gain of 10$^{9}$ A/V. Light collection from the STM tunnel junction is done with a silver off-axis parabolic mirror that has a focal length of 33.8 mm and a diameter of 1 inch with an incident angle of 60 degrees and an estimated numerical aperture of 0.35. Optical spectra were collected through a fiber collimator using a Teledyne Princeton Instruments SpectraPro HRS 300 spectrograph, equipped with an LN-cooled PyLoN 400BR eXcelon detector with an estimated spectral resolution of $\sim$4 nm and $\sim$2 nm after diffracted from a grating of 150 l/mm (\figref{2}c, \figref{3}b/c, \figref{4}) and 600 l/mm (\figref{1}b), respectively. A Ag wire tip was used for all STM and STML measurements and was prepared by repeated indentations into the metallic substrate and voltage pulses. The optical response of the STM tip was optimized by repeated indentation until a plasmonic emission spectrum measured on the bare Ag surface at +2.5\,V on Ag(111) or -2.5\,V on Ag(110) overlapped with the spectral emission of the \TBP\ molecule. Each STML emission spectrum is normalized by the tip-specific plasmonic response present at the time of measurement to account for intensity variations of the junction response. The hyper-resolved fluorescence mapping in \figref{4}b-d was performed at constant height with a drift correction procedure. Drift correction was performed in constant-current conditions on the NaCl surface with a setpoint current of 10\,pA and sample bias of 1.4\,V (\To) and 1.0\,V (\Tz). The fluorescence maps were recorded with a constant tip-height offset relative to the NaCl surface of $\Delta$Z = 450\,pm (\To) and 437\,pm (\Tz). The fluorescence maps have a pixel-to-pixel distance of 50\,pm and are restricted to locations near the molecule where excitation currents were greater than 0.2\,$\%$ of the maximum current ($\sim$300-500\,fA).

\subsection{Time-dependent Density Functional Theory}
Density functional theory (DFT) and time-dependent DFT (TD-DFT) calculations were performed in gas-phase using B3LYP functional and 6-31G$^\ast$ basis set within the Gaussian16 software package~\cite{Gaussian}. TD-DFT used the Tamm-Dancoff approximation (TDA) and the emission spectra were simulated for a Lorentzian lineshape with broadening of 40 cm$^{-1}$ half width at half maximum. All calculations were performed in gas phase.

\clearpage
\section{Supporting Information}
The Supporting Information is available free of charge:

Voltage dependent STML spectra; detailed description of polaron and rate equation model; decoupling properties of NaCl on Ag(111) and Ag(110) (PDF)

\section{Associated Content}
A pre-print version of this manuscript has been submitted to arXiv:
Eve Ammerman; Nils Krane; Bruno Schuler; Tunable Luminescence From a Single Free-Base Porphyrin Molecule By Controlled Access to Optically Active States (2025) 2511.00630, arxiv.org, https://arxiv.org/abs/2511.00630 (accessed January 23, 2026).

\section*{Data availability}
The data that support the findings of this study are available from the corresponding author upon request.

\section*{Acknowledgments}
The authors would like to acknowledge Dr. Andres Ortega-Guerrero for his expert guidance on performing DFT and TD-DFT calculations. We would also acknowledge fruitful discussions with Dr. Oliver Gröning and Prof. Roman Fasel. This research was funded by the Empa Young Scientist Fellowship and the European Research Council (ERC) under the European Union’s Horizon 2020 research and innovation program (Grant agreement No. 948243). EA and NK appreciate financial support from the Werner Siemens Foundation (CarboQuant). For the purpose of Open Access, the author has applied a CC BY public copyright license to any Author Accepted Manuscript version arising from this submission.

\section{Author contributions}
EA and BS designed the experiments while measurements were performed by EA. NK was responsible for DFT/TD-DFT calculations, both EA and NK contributed to STML modeling, and all authors contributed to writing of the manuscript.

\section*{Competing interests}
The authors declare no competing interests.

\clearpage
\bibliography{references}

\clearpage 

\section{Figures}

\clearpage 

\clearpage 

\clearpage 

\end{document}


\title{Supporting Information: Tunable Luminescence From a Single Free-Base Porphyrin Molecule By Controlled Access to Optically Active States}

\author{Eve Ammerman\,\orcidlink{0000-0003-3588-2245}}
\email{eve.ammerman@empa.ch}
\affiliation{nanotech@surfaces Laboratory, Empa -- Swiss Federal Laboratories for Materials Science and Technology, D\"ubendorf 8600, Switzerland}

\author{Nils Krane\,\orcidlink{0000-0002-8273-6390}}
\affiliation{nanotech@surfaces Laboratory, Empa -- Swiss Federal Laboratories for Materials Science and Technology, D\"ubendorf 8600, Switzerland}

\author{Bruno Schuler\,\orcidlink{0000-0002-9641-0340}}
\affiliation{nanotech@surfaces Laboratory, Empa -- Swiss Federal Laboratories for Materials Science and Technology, D\"ubendorf 8600, Switzerland}


\date{\today}
\pacs{}
\maketitle

\tableofcontents
\newpage

\clearpage

\section{Polaron Model}
The observed d$I$/d$V$ peaks of molecules on NaCl are known to be, both, broadened and shifted to higher absolute energy, due to the strong electron-phonon coupling of the underlying NaCl~\cite{Repp2005,Fatayer2018}. To extract the zero-phonon ion resonance energy, we employ a simple polaron model, as described by Vasilev et al.~\cite{vasilev2024}: 
\begin{align}
    \label{eq:polaron}
    S(E_{fi},Q_{fi},V) &= \frac{1}{2\pi}Re\left \{ 
    \int e^{-ieVt/\hbar} \
    e^{\mathcal{F}(E_{fi},Q_{fi})}\ \mathrm{d}t
    \right \}, \quad \mathrm{with}\\
    \mathcal{F}(E_{fi},Q_{fi}) &= -iQ_{fi}E_{fi}t/\hbar + \int_0^\infty  J(\Omega)\ 
    e^{-iQ_{fi}\Omega t / \hbar}\ \mathrm{d}\Omega,
\end{align}
where $E_{fi}$ is the energy difference between final and initial state and $Q_{fi}$ a unitless charge number, being $+1$ or $-1$ for hole or electron attachment, respectively. The electron-phonon coupling strength of the NaCl is included in the rectangular function $J(\Omega)$, which we assume to have a constant value $\eta$ between $\Omega_\mathrm{min} = 18$\,meV and $\Omega_\mathrm{max} = 31$\,meV and is zero otherwise~\cite{vasilev2024}.

\begin{figure}[htbp]
    \includegraphics[width=\textwidth]{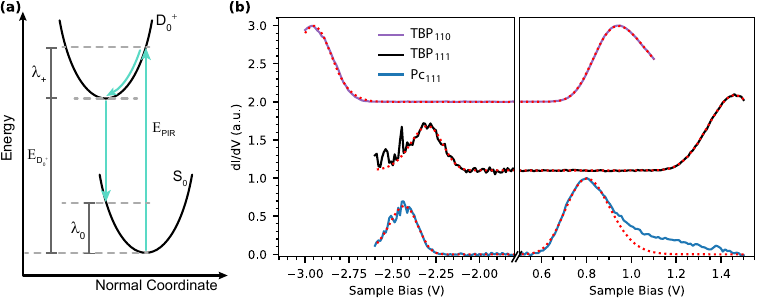}
    \caption{\textbf{Polaron Modeling of d$I$/d$V$ measurements.}
        (a) Sketch of the free-energy curves of a neutral and positively charged molecule on NaCl. The relaxation energy for ionization and neutralization are labeled $\lambda_+$ and $\lambda_0$, respectively.
        (b)
        Experimental d$I$/d$V$ spectra taken on TBP on NaCl(4ML)/Ag(110) (purple) and NaCl(3ML)/Ag(111) (brown) as well as Pc on NaCl(3ML)/Ag(111). Dotted red lines correspond to the model fit with respect to the corresponding d$I$/d$V$ data.  
        }
    \label{fig:polaron}
\end{figure}

The parameters $E_{fi}$ and $\lambda_\pm$ are used as fit parameters for the polaron model to the d$I$/d$V$ spectra. Here, $E_{fi}$ corresponds to the zero-phonon ion resonance energy (\textit{cf.} $E_\mathrm{PIR}$ in figure \ref{fig:polaron}a) and $\lambda_\pm$ is proportional to the electron-phonon coupling strength $\eta$:
\begin{equation}
    \lambda = \int_0^\infty J(\Omega)\Omega\ \mathrm{d}\Omega = 
    \frac{\eta}{2} \left ( \Omega_\mathrm{max}^2 - \Omega_\mathrm{min}^2\right )
\end{equation}
Fitting was performed with a voltage scale of 0.9 to account for voltage drop of $\alpha_\mathrm{P}=10\%$ across the NaCl. In \figref{fig:polaron} the x-axis of the fit data is rescaled to match the experimental data.

\begin{table}[htbp]
    \setlength{\tabcolsep}{10pt}
    \centering
    \caption{\textbf{Fitted parameters obtained from polaron model ($\alpha_\mathrm{P} = 10\%$).}}
    \begin{tabular}{cccc}
        \toprule
        \textbf{Parameter} & \textbf{\To} & \textbf{\Po} & \textbf{\Tz} \\
        \midrule
        $E_{D_0^+}$(meV) & 1812\,$\pm$\,14 &  1984\,$\pm$\,5  & 2248\,$\pm$\,5 \\ 
        $\lambda_+$(meV) & 268\,$\pm$\,11 & 216\,$\pm$\,6  & 303\,$\pm$\,7 \\
        $E_{D_0^-}$(meV) & 937\,$\pm$\,8 &  422\,$\pm$\,6  & 544\,$\pm$\,2 \\ 
        $\lambda_-$(meV) & 310\,$\pm$\,11 & 311\,$\pm$\,7  & 297\,$\pm$\,3 \\
        \bottomrule
    \end{tabular}
    \label{tab:fitParameterPolaron}
\end{table}

\clearpage

\section{Voltage Dependent STML Intensites}

\begin{figure}[htbp]
    \centering
    \includegraphics[width=0.9\linewidth]{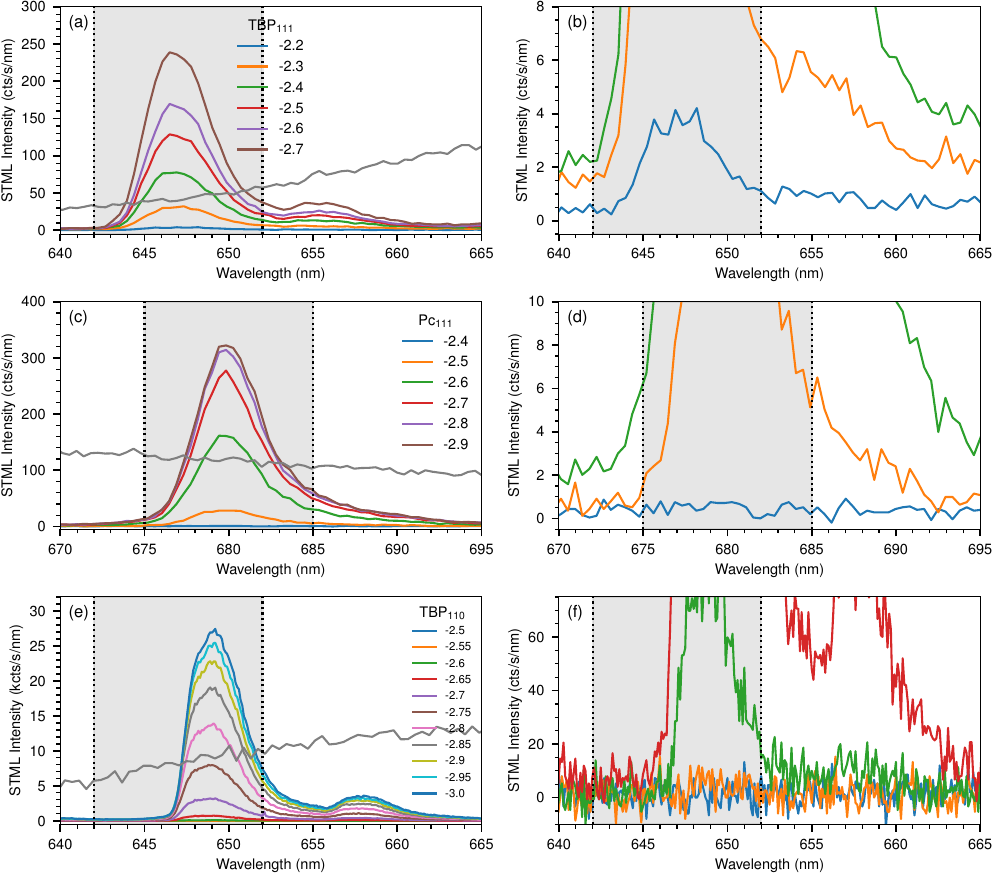}
    \caption{STML spectra taken on \TBP/NaCl/Ag(111), \Pc/NaCl/Ag(111) and \TBP/NaCl/Ag(110). The right column shows a zoom in onto the low intensity spectra taken at low bias voltages. The gray area marks the integration area and the grey line represents the reference plasmon spectrum in arbitrary units.}
    \label{fig:raw_stml}
\end{figure}

\clearpage

\section{Rate Equation Model for Voltage Dependent Emission}
The voltage dependence of the emission rate was simulated with a simple rate equation model, similar to the one discussed in Ref.~\citenum{kaiser2025}. Here we consider only three states (see Fig. 2a of the main text), namely $S_0$, $S_1$ and $D_0^+$. The energy $E_{S_1}=1.915$\,eV of the excitonic state is given by the experimentally measured photon energy of the primary emission line and from the polaron fit we extracted the energy $E_{D_0^+}$

The voltage dependent transition rates between the neutral and the charged state are given by integration of the polaron spectral profile as described in equation \eqref{eq:polaron}:
\begin{align*}
    \Gamma^{t/s}_{S_i\rightarrow D_0^+}(V_\mathrm{eff}) &= \frac{2\hat{V}_{t/s}}{h}\int_{V_\mathrm{eff}}^\infty  S(E_{D_0^+}-E_{S_i},+1,V')\ \mathrm{d}V' \\
    \Gamma^{t/s}_{D_0^+\rightarrow S_i}(V_\mathrm{eff}) &= \frac{\hat{V}_{t/s}}{h} \int_{-\infty}^{V_\mathrm{eff}}  S(E_{S_i}-E_{D_0^+},-1,V')\ \mathrm{d}V',
\end{align*}
with $\hat{V}_{t/s}$ being the coupling strength of the molecule to tip and surface. The effective bias voltage $V_\mathrm{eff}$ between molecule and corresponding lead is given by the voltage drop $\alpha_\mathrm{RE}\approx0.1$: 
\begin{align}
    V^t_\mathrm{eff} &= (1-\alpha_\mathrm{RE})V_b  \\
    V^s_\mathrm{eff} &= -\alpha_\mathrm{RE} V_b.
\end{align}
The additional factor of two for the singlet to doublet transition $S_i\rightarrow D_0^+$ yields from the multiplicities of the involved states and is given by the Clebsch-Gordan coefficients $\Upsilon_{fi}$. 
As an approximation, we assume the relaxation energy $\lambda$ to be the same for all transitions between the positive ion resonance $D_0^+$ and the charge neutral states $S_0$ and $S_1$:
\begin{equation}
  \lambda = \lambda_+^0 = \lambda_0^0 = \lambda_+^1 = \lambda_0^1  
\end{equation}

In the bias range of interest we further assume some transition rates to be constant:
\begin{align}
    \Gamma^s_{S_0\rightarrow D_0^+} &= 0 \\
    \Gamma^s_{D_0^+\rightarrow S_0} &= 2.5 \hat{V_s}/h \\
    \Gamma^t_{D_0^+\rightarrow S_0} &= \Gamma^t_{D_0^+\rightarrow S_1} = 0 \\
    \Gamma^t_{S_1\rightarrow D_0^+} &= 2 \hat{V_t}/h 
\end{align}
We estimated the coupling of the molecule to the substrate to be $\hat{V}_s=4$\,$\mu$eV (see below) for all orbitals and used the coupling to the tip $\hat{V}_t$ as a fitting parameter.
In this simplified model we do not directly include $T_1$, since experimentally there is no indication of triplet pumping \cite{kaiser2025}. However, the presence of the triplet state causes an increase in the dark current as it opens another channel to neutralize the molecule via sample from the $D_0^+$ state into $T_1$ and then decay non-radiatively back into the ground state $S_0$. To capture this effect, the transition rate $\Gamma^s_{D_0^+\rightarrow S_0}$ is multiplied by a factor of 2.5, which is the sum of the Clebsh-Gordan coefficients $\Upsilon_{D_0^+\rightarrow S_0} = 1$ and $\Upsilon_{D_0^+\rightarrow T_1} = 1.5$.
For the radiative emission $\Gamma^\mathrm{rad}_{S_1\rightarrow S_0} = \hat{V}_\mathrm{rad}/h$ the lifetime was estimated from the width of the emission peak to be in the order of $\hat{V}_\mathrm{rad} = 1$\,meV.

The occupation probability $N_i$ of the three considered states $S_0$, $S_1$ and $D_0^+$ is calculated using a differential rate equation for the steady state condition
\begin{equation}
    0 = \sum_{j} \Gamma_{j\rightarrow i}(V_b) N_j - \Gamma_{i\rightarrow j}(V_b) N_i
\end{equation}
for given bias voltage $V_b$.
The net current flowing through the molecule is then defined by 
\begin{equation}
    I_\mathrm{mol}(V) = -e^- (\Gamma^t_{S_0\rightarrow D_0^+}N_{S_0} + \Gamma^t_{S_1\rightarrow D_0^+}N_{S_1}).    
\end{equation}
The number of photons emitted is given by the probability of occupation of \ESa\ and the radiative emission rate:
\begin{equation}
    J^0 _\mathrm{ph}(V) = \Gamma^\mathrm{rad}_{S_1\rightarrow S_0}N_{S_1}.
\end{equation}
To account for the experimental detection efficiency, a prefactor $0<A<1$ is introduced to obtain the experimental photon counts:
\begin{equation}
    J _\mathrm{ph}(V) = A\ J^0_\mathrm{ph}(V) =A\ \Gamma^\mathrm{rad}_{S_1\rightarrow S_0}N_{S_1}
\end{equation}
with an expectation value of about $A\sim 10^{-4}$~\cite{Kuhnke2017}.

\begin{figure}[htbp]
\includegraphics[width=0.9\textwidth]{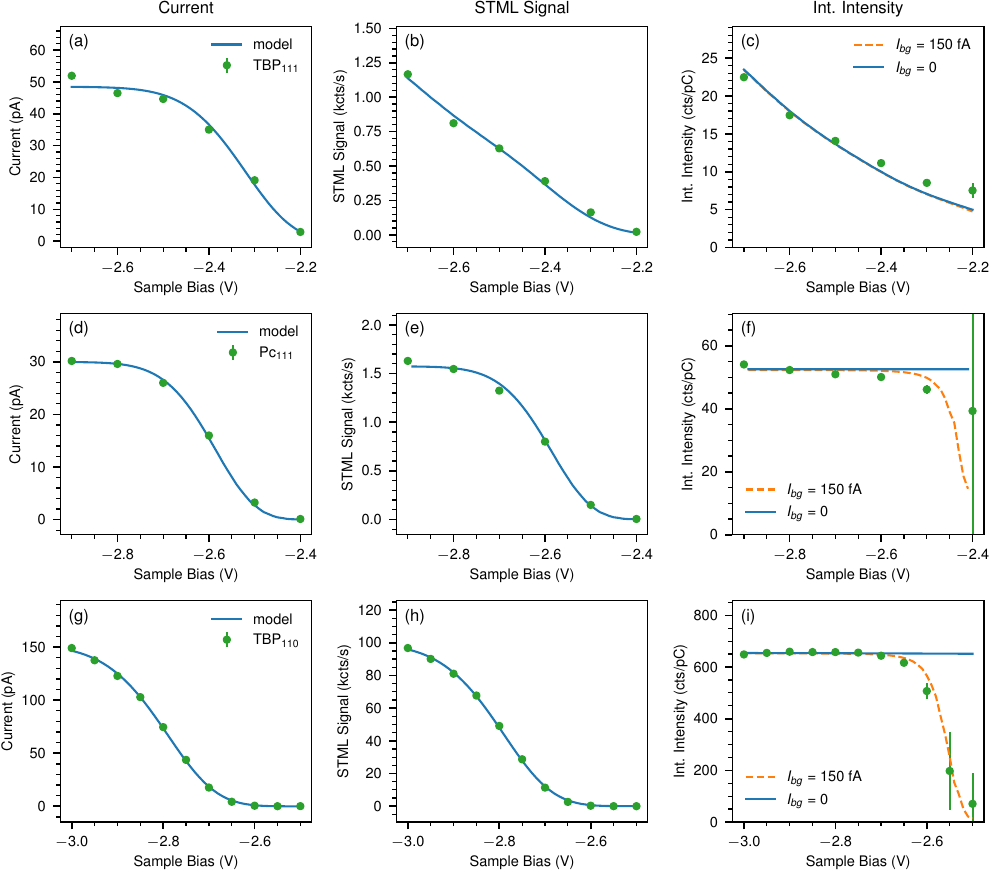}
\caption{\label{fig:ModelFit}\textbf{Fitting of STML data.} 
    Bias dependency of the experimental current and STML intensity of the dominant emission line, as well as the current normalized integrated intensity for \To\ (a-c), \Po\ (d-f) and \Tz\ (e-h). The blue solid lines are  fit of the rate equation model to the experimental data, and the orange dashed line represents current normalized STML intensity, assuming the presence of an additional background current of 150\,fA.
}
\end{figure}

The modeled current and light intensity are fitted to the experimental data, using tip coupling $\hat{V}_t$, voltage drop $\alpha_\mathrm{RE}$ and photon detection efficiency $A$ as fit parameters. The resulting fits are shown in \figref{fig:ModelFit} and the corresponding parameter listed in table~\ref{tab:fitParameter}.

It should be noted that the calculated STML intensity for \To\ is very sensitive to the voltage drop $\alpha_\mathrm{RE}$, because the system is right at the onset regime of emission and slight changes in $\alpha_\mathrm{RE}$ cause an exponential increase or decrease in emission intensity. Accordingly, the obtained error value for $\alpha_\mathrm{RE}$ from the fit is less than $1\cdot10^{-4}$ percent points. However, the value of $\alpha_\mathrm{RE}$ itself also depends on the voltage drop $\alpha_\mathrm{P}$ used for the polaron fit, effectively increasing the actual error margin of $\alpha_\mathrm{RE}$ (see section below).

For the normalized STML intensity one also needs to consider the effect of an additional background current flowing directly between the tip and the underlying substrate. The total current is given by the sum of the current through the molecule ($I_\mathrm{mol}$) and the additional background current ($I_\mathrm{bg}$):
\begin{align}
    I_\mathrm{tot} = I_\mathrm{mol} + I_\mathrm{bg},
\end{align}
with $I_\mathrm{tot}$ corresponding to the current measured in the experiment. In particular, at the onset of the PIR, these two currents become comparable in magnitude ($I_\mathrm{mol} \approx I_\mathrm{bg}$). In especially in the case of \To\ and \Po\ this causes a drop in the normalized emission intensity
\begin{equation}
    j_\mathrm{ph} = \frac{J_\mathrm{ph}}{I_\mathrm{tot}}
    =  \frac{J_\mathrm{ph}}{I_\mathrm{mol} + I_\mathrm{bg}},
\end{equation}
although $J_\mathrm{ph}/I_\mathrm{mol}$ is expected to be constant in the bias range of interest.
To account for this effect in the simulation, we added an additional background current of $I_\mathrm{bg}=150$\,fA to the calculated current $I_\mathrm{mol}$ to reproduce the drop in $j_\mathrm{opt}$, as shown in Fig.~\ref{fig:ModelFit}.

\begin{table}[htbp]
    \setlength{\tabcolsep}{10pt}
    \centering
    \caption{\textbf{Parameters used to model the STML intensity.}
    $E_{D_0^+}$ and $\lambda^+$ were obtained from the polaron model using $\alpha_\mathrm{P}=10\%$. $A$, $\alpha_\mathrm{RE}$ and $\hat{V}_t$ were used as fitting parameters for the rate equation model.}
    \begin{tabular}{cccc}
        \toprule
        \textbf{Parameter} & \textbf{\To} & \textbf{\Po} & \textbf{\Tz} \\
        \midrule
        $E_{S_0}$ (eV) & \multicolumn{3}{c}{0} \\
        $E_{S_1}$ (eV) & 1.915 & 1.810 & 1.915 \\
        $E_{D_0^+}$ (eV) & 1.812 &  2.123 & 2.248 \\ 
        $\lambda^+$ (meV) & 268 & 186 & 303 \\
        $A$  ($\times 10^{-4}$)& $1.08\pm0.04$ & $0.30\pm0.01$ &  $3.70\pm0.03$ \\
        $\alpha_\mathrm{RE}$ (\%) & 10.4 & $10.8\pm0.2$ & $ 9.0\pm0.1$ \\
        $\hat{V}_t$ ($\mu$eV) & $0.70\pm0.02$ & $0.41\pm0.01$ & $2.73\pm0.03$ \\
        $\hat{V}_s$ ($\mu$eV) & \multicolumn{3}{c}{4} \\
        \bottomrule
    \end{tabular}
    \label{tab:fitParameter}
\end{table}

\begin{figure}[htbp]
    \centering
    \includegraphics[width=0.6\linewidth]{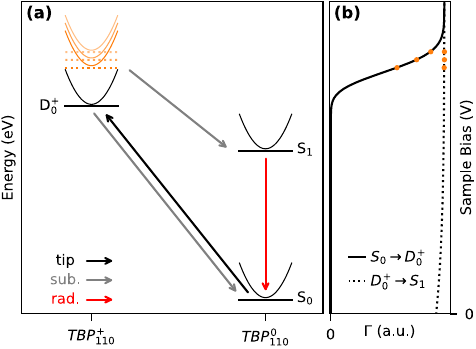}
    \caption{(a) Many-body diagram describing the electronic states and transitions of a \Tz\ molecule during STML experiments, analogous to Fig.~2a,b in the main text. Black and grey arrows connecting different \Tz\ charge states represent electron transfer involving the tip and substrate while the red arrow represents radiative relaxation. The excess relaxation energy needed for charge transport is depicted as a parabola above each state. A voltage dependent shift in \PIR\ energy, $\Delta E(V)$, is present during experiments and indicated by a series of orange dashed lines. (b) Voltage dependent \GS\ $\rightarrow$ \PIR\ and \PIR\ $\rightarrow$ \ESa\ transition rates. Orange circles correlate with $\Delta E(V)$ shown in (a).}
    \label{fig:MBD_TBP110}
\end{figure}
\clearpage

\section{Role of Voltage Drop in Simulation}
The voltage drop over the NaCl layer is a crucial parameter to model the system and simulate the light emission from the junction. In the following we will be discussing the effect of different assumed voltage drops $\alpha_\mathrm{P}$ in the polaron fit.

\begin{figure}[htbp]
    \centering
    \includegraphics[width=0.6\linewidth]{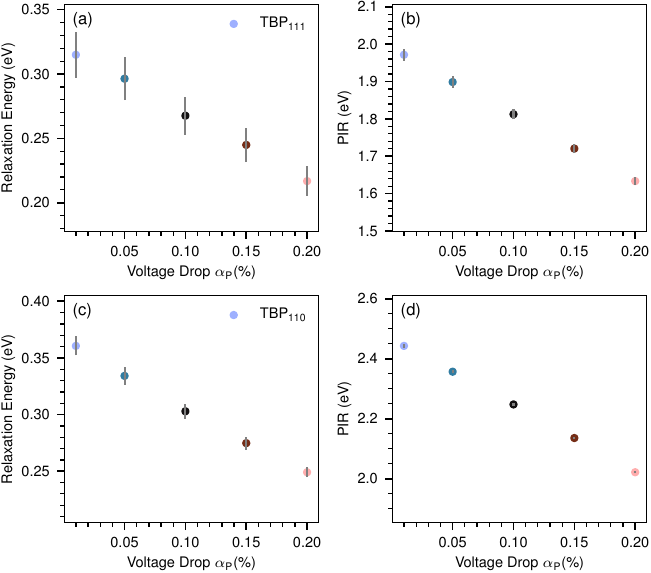}
    \caption{Dependency of relaxation energy $\lambda^+$ and ion resonance energy $E_{D_0^+}$ on assumed voltage drop $\alpha_\mathrm{P}$, obtained by polaron fit.}
    \label{fig:VD_fitparams_polaron}
\end{figure}

We fitted the experimental d$I$/d$V$ spectra assuming five different values for $\alpha_\mathrm{P}$, ranging between 1\% and 20\%. The obtained best fit parameters $\lambda^+$ and $E_{D_0^+}$ for each value of $\alpha_\mathrm{P}$ are shown in Fig.~\ref{fig:VD_fitparams_polaron} for both \To\ (top row) and \Tz\ (bottom row). With increasing voltage drop, we observe an almost linear reduction in relaxation energy $\lambda^+$, as well as in the ion resonance energy $E_{D_0^+}$, with good fits to experiment in all cases. In the strong electron-phonon coupling regime, with relaxation energies much larger than the phonon energies ($\lambda^+ \gg \hbar\Omega$), an increase (decrease) in relaxation energy will linearly widen (narrow) the ion resonance peak and shift it to higher (lower) energies, while maintaining the approximate gaussian line shape. For this reason, the experimental spectra can be fitted well with any voltage drop in the voltage drop range considered here.

\begin{figure}[htbp]
    \centering
    \includegraphics[width=0.9\linewidth]{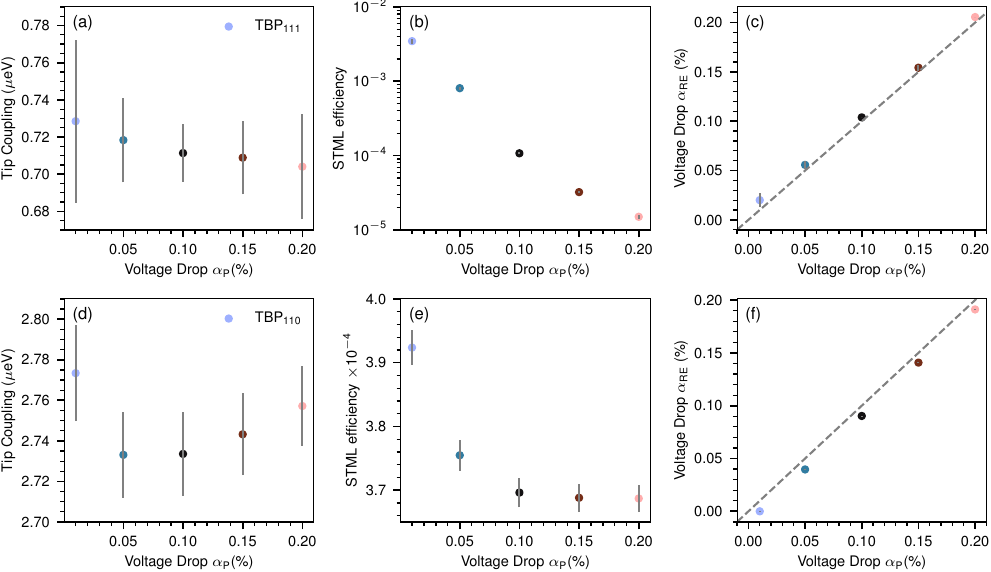}
    \caption{Dependency of tip coupling strength $\hat{V}_t$, STML efficiency $A$ and rate equation voltage drop $\alpha_\mathrm{RE}$ on the assumed polaron voltage drop $\alpha_\mathrm{P}$, obtained by rate equation fit.}
    \label{fig:VD_fitparams_re}
\end{figure}

\begin{figure}[htbp]
    \centering
    \includegraphics[width=0.9\linewidth]{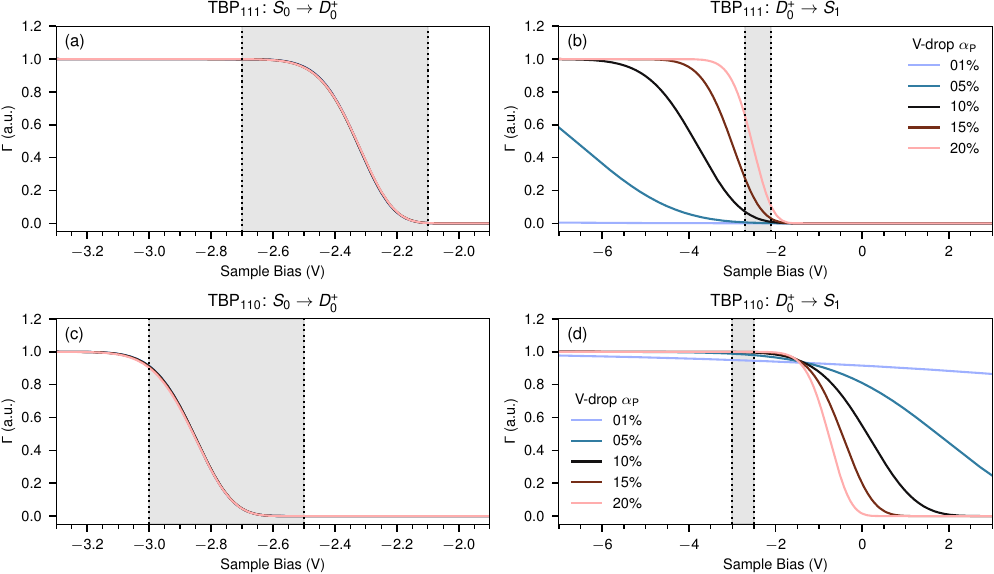}
    \caption{Dependency of the transition rates on the assumed polaron voltage drop $\alpha_\mathrm{P}$, obtained by rate equation fit. The gray areas mark the voltage range of the experiment.}
    \label{fig:VD_transition_rates}
\end{figure}

In the next step, the obtained values for the relaxation and ion resonance energy were used in the rate equation model, to reproduce the experimental data sets as described in the section above. Fig.~\ref{fig:VD_fitparams_re} shows the best fit parameters in dependence of the polaron voltage drop $\alpha_\mathrm{P}$, again for \To\ and \Tz. In both cases, the tip coupling strength does not depend strongly on $\alpha_\mathrm{P}$, which to be expected, as this parameter mainly scales the amplitude of the current.
On the other hand, the STML efficiency $A$ depends almost exponentially on $\alpha_\mathrm{P}$ for \To, spanning several orders of magnitude. In contrast, it is nearly constant for \Tz. This effect can be rationalized, by a closer look to the voltage dependent transition rates $\Gamma^t_{S_0\rightarrow D_0^+}$ and $\Gamma^s_{D_0^+\rightarrow S_1}$ for different voltage drops, as depicted in Fig.~\ref{fig:VD_transition_rates}. The tip mediated rate $\Gamma^t_{S_0\rightarrow D_0^+}$ is the same for all $\alpha_\mathrm{P}$ (Fig.~\ref{fig:VD_transition_rates}a,c), since it represents the main contribution to the tunneling current and is accordingly fitted to experiment. In contrast, the sample mediated rate $\Gamma^s_{D_0^+\rightarrow S_1}$ depends strongly on the initially assumed voltage drop (Fig.~\ref{fig:VD_transition_rates}b). The gray areas indicate the voltage range of the experimental data set. For \To\ the voltage range is just at the exponential onset of the transition rate, in especially for small values of $\alpha_\mathrm{P}$. Accordingly, $\Gamma^s_{D_0^+\rightarrow S_1}$ increases exponentially with increasing $\alpha_\mathrm{P}$ for a given voltage in that range. In order to fit the simulated light emission $J_\mathrm{ph} = A\ J^0_\mathrm{ph}$ to experimental data, the STML efficiency $A$ has to decrease exponentially, as we observed. For \Tz\ the voltage range (gray area in Fig.~\ref{fig:VD_transition_rates}d) is already at the plateau of $\Gamma^s_{D_0^+\rightarrow S_1}$, hence the near constant values of $A$.

The obtained values for $\alpha_\mathrm{RE}$ are shown in Fig.~\ref{fig:VD_fitparams_re}c,f. As expected, they show a near linear behavior, with $\alpha_\mathrm{RE}$ being within one percent point around $\alpha_\mathrm{P}$. The slight deviation of the two voltage drop parameters are reasoned by the tip shape and position dependency of the voltage drop, which can lead to slight difference between data sets.

\begin{figure}[htbp]
    \centering
    \includegraphics[width=0.9\linewidth]{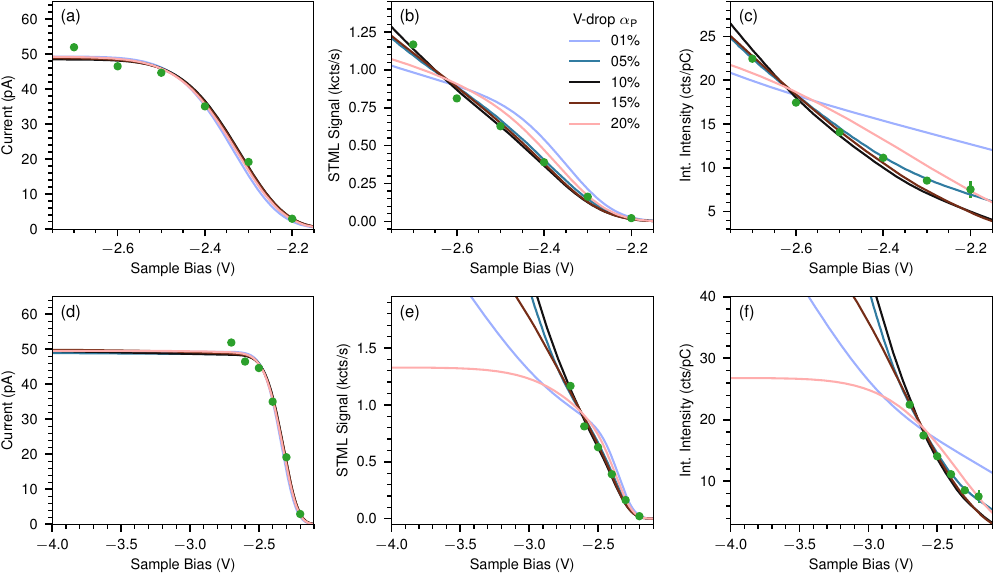}
    \caption{STML fits for \To\ for several assumed polaron voltage drop $\alpha_\mathrm{P}$. In the bottom row, the simulated curves are extrapolated to higher absolute bias voltages.}
    \label{fig:Porph111_VD_STMLfits}
\end{figure}

\begin{figure}[htbp]
    \centering
    \includegraphics[width=0.9\linewidth]{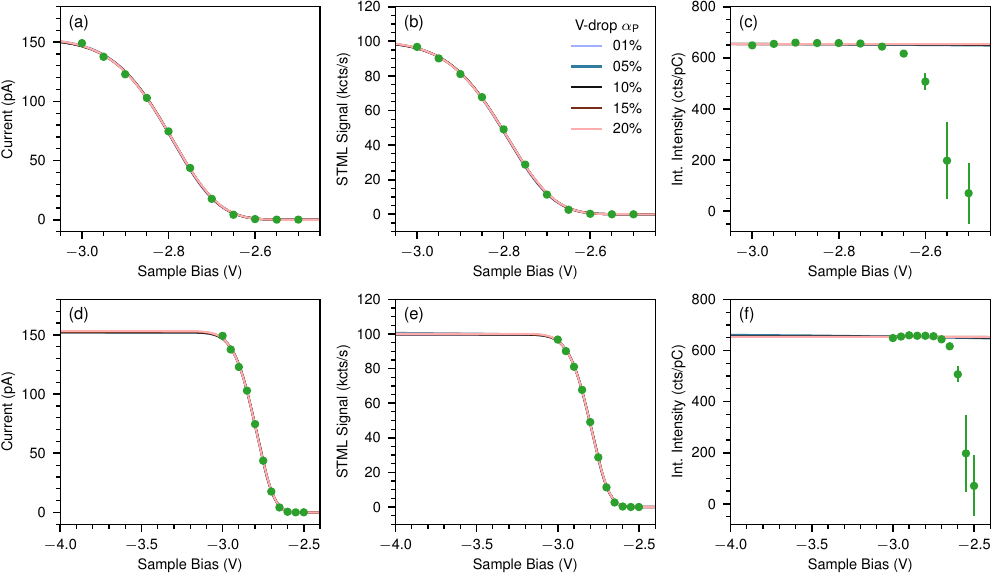}
    \caption{STML fits for \Tz\ for several assumed polaron voltage drop $\alpha_\mathrm{P}$. In the bottom row, the simulated curves are extrapolated to higher absolute bias voltages.}
    \label{fig:Porph110_VD_STMLfits}
\end{figure}

The effect of the assumed voltage drop on the simulated STM current and STML intensity is shown in Fig.~\ref{fig:Porph111_VD_STMLfits} for \To\ and Fig.~\ref{fig:Porph110_VD_STMLfits} for \Tz. In the first case, the STML intensity $J_\mathrm{ph}$ can be fitted decently for most voltage drops, except for $\alpha_\mathrm{P}=20\%$, which saturates too fast and cannot capture the experimental behavior at higher bias voltages. The deviation of $J_\mathrm{ph}$ for $\alpha_\mathrm{P}=1\%$ is caused by a shoulder in the transition rate $\Gamma^s_{D_0^+\rightarrow S_1}$, due to discernable vibronic states on the low energy flank of the polaron peak (see Fig.~\ref{fig:Porph111_VD_STMLfits}e). These shoulders depend strongly on the electron-phonon coupling function $J(\Omega)$ (see above) and are amplified by the very small voltage of 1\%, but should not be considered an exact modeling of the transition rate, since $J(\Omega)$ is approximated by a simple rectangular function. Nevertheless, we can conclude from these fits that the voltage drop for \To\ has to be smaller than 20\% to fit with the experiment.
Another estimate can be done by consideration of the STML efficiency $A$. A typical estimation of the STML efficiency is $A\sim10^{-4}$~\cite{Kuhnke2017}, which would match with a voltage drop between 5\% and 15\% (see Fig.~\ref{fig:VD_fitparams_re}b).

For \Tz, the assumed voltage drop has little effect on the STML fit, therefore we cannot deduce how strong the voltage drop is in the experiment. Since the transition $\Gamma^s_{D_0^+\rightarrow S_1}$ is always fully allowed, the STML efficiency parameter $A$ is nearly constant as well, with $A\sim4\cdot10^{-4}$.

In accordance with literature, we decided to assume a voltage drop of 10\% for both substrates. The experimental value is expected to be close within a few percent points and the precise value of $\alpha$ does not change significantly the observed voltage dependent behavior of the STML intensity.

\clearpage

\section{NaCl Decoupling on Ag(111) and Ag(110)}

\begin{figure}
    \centering
    \includegraphics[width=0.6\linewidth]{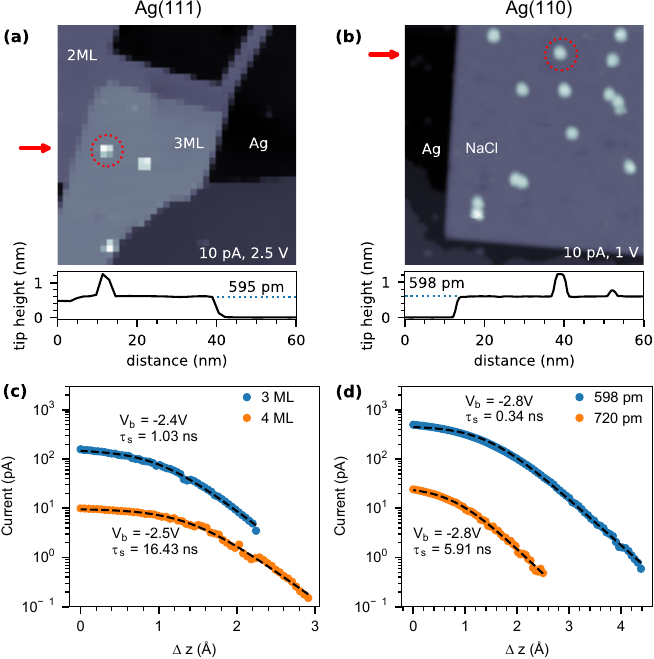}
    \caption{Overview images of \TBP\ molecules adsorbed on few-layer NaCl films of comparable topographic heights grown on (a) Ag(111) and (b) Ag(110). STM topographic height profiles are shown along the bottom axis of the images with line profile locations indicated by red arrows. (c) I(z) approach curves measured on \TBP\ molecules adsorbed on 3 ML NaCl (blue dots) and 4 ML NaCl (orange dots) grown on Ag(111). The 3 ML curve is measured on the molecule highlighted with a red dashed circle. (d) I(z) approach curves measured on \TBP\ molecules adsorbed on NaCl islands grown on Ag(110) with apparent heights comparable to those shown in (c). Black dashed lines show fits to the I(z) data which were used to extract charge state lifetimes, $\tau_{s}$, indicated in the figures. Each measurement was recorded with a sample bias, $V_{b}$, shown alongside the corresponding I(z) curve. The molecule highlighted in (a) was used to record the STML map shown in Fig. 4a while the highlighted molecule in (b) was used in Fig. 3c.}
    \label{fig:approach_curves}
\end{figure}

The charge state lifetime of \TBP\ on the different NaCl layers was determined by fitting a simple rate model to $I(z)$ curves, using the function~\cite{Kaiser2023}:
\begin{equation}
    I(z) = \frac{e^-}{\tau_s + \tau^0_t \ e^{2\kappa \Delta z}},
\end{equation}
with $e^-$ being the elementary charge, $\tau_s = 1/\Gamma_s$ the lifetime of the charge state, $\tau^0_t = 1/\Gamma^0_t$ the inverse tunneling rate from tip to molecule at relative tip height $\Delta z=0$ and $\kappa$ being the decay constant.

\section{Simulation of Fluorescence Maps}
\begin{figure}[htbp]
    \centering
    \includegraphics[width=0.9\linewidth]{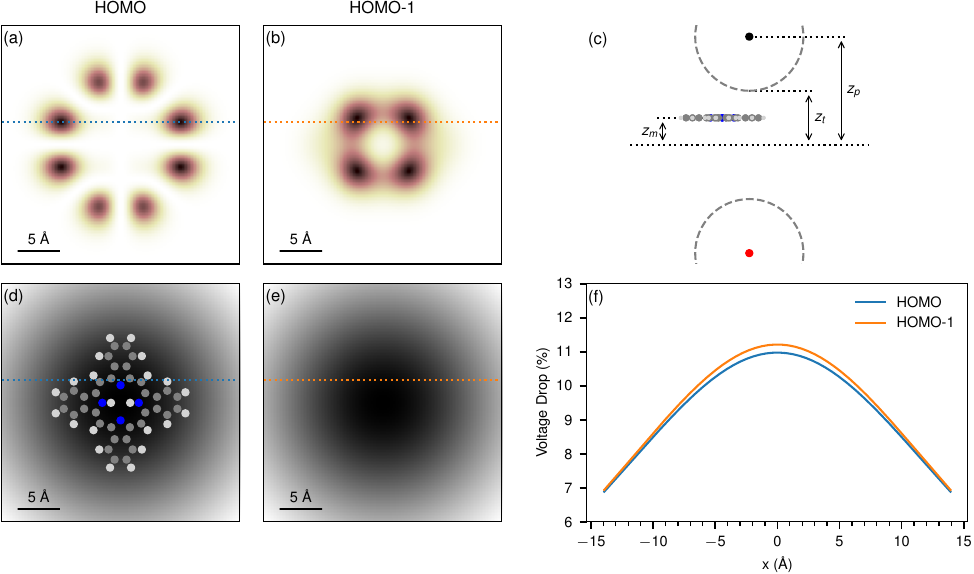}
    \caption{\TBP\ in the STM junction.
    (a,b) Spatial dependence of the tip--molecule coupling to the HOMO and the HOMO-1, respectively, for a tip--molecule distance of 10\,\AA.
    (c) Schematic of the junction model, represented by two point charges of opposite sign. The molecule is positioned at a height $z_m$ above the surface plane and the tip (grey dashed line) is assumed to have a radius of $R=z_p-z_t$.
    (d,e) Spatial dependence of the effective voltage drop for the HOMO and the HOMO-1, respectively.
    (f) Line cuts along the dashed lines in (a,b) and (d,e), showing the effective voltage drop $\alpha$.
    }
    \label{fig:simMaps_VD}
\end{figure}

The hyper-resolved flourescence maps were simulated using the rate equation and polaron model described above. Here, we consider tunneling into the HOMO and HOMO-1 as possible excitation pathways. The spatial dependence of the tip--molecule coupling strength $\hat{V}_t$ is obtained from DFT calculations by extrapolating the single-particle wave function $\Psi_i(\mathbf{r})$ of molecular orbital $i$ into the vacuum region:
\begin{equation}
    \hat{V}^i_t(\mathbf{r_t}) = \hat{V}_t^0 \frac{\left | \Psi_i(\mathbf{r_t})\right|^2 }{
    \mathrm{max} \left ( \left | \Psi_i(\mathbf{r_t})\right|^2 \right )},
\end{equation}
where $\hat{V}_t^0$ denotes the maximum tip–molecule coupling strength, corresponding to the value obtained from fits with the rate-equation model. Figure~\ref{fig:simMaps_VD}a,b show the spatial dependence of the tip coupling for the HOMO and HOMO–1, respectively, extrapolated to a tip height of 10\,\AA.

\begin{figure}[htbp]
    \centering
    \includegraphics[width=0.9\linewidth]{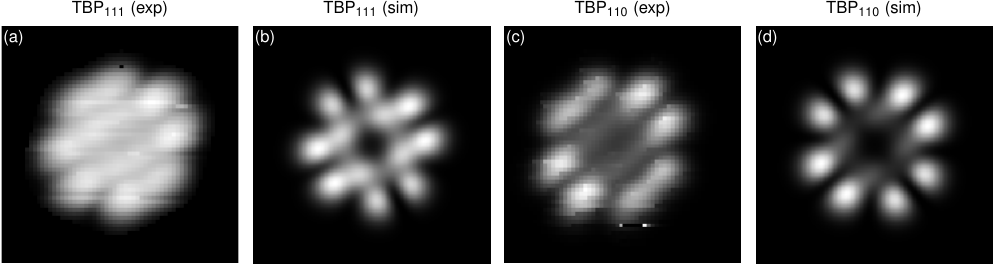}
    \caption{Experimental current maps, extracted from the spatially resolved fluorescence maps, and spatial dependency of the tip coupling strength, according to eq.~\eqref{eq:tip_coupling}, for \To\ (a,b) and \Tz\ (c,d).
    }
    \label{fig:simMaps_current}
\end{figure}

For the effective tip--molecule coupling, we consider tunneling contributions into both the HOMO and the HOMO-1:
\begin{equation}
    \hat{V}_t(\mathbf{r_t}) = \hat{V}^\mathrm{HOMO}_t(\mathbf{r_t}) +
    \beta\  \hat{V}^\mathrm{HOMO-1}_t(\mathbf{r_t})
    \label{eq:tip_coupling}
\end{equation}
Excitation via the HOMO–1 orbital would populate the higher-energy state $D_1^+$. However, due to the long lifetime of the charged state (on the order of several hundred picoseconds), we can assume that the molecule relaxes to \PIR\ before neutralizing into \ESa. The mixing ratio $\beta$ depends on the experimental bias voltage. In the simulations, we use $\beta_{111}=1.0$ for \To\ and $\beta_{110}=0.4$ for \Tz. The comparison between experimental and simulated current maps is shown in Figure~\ref{fig:simMaps_current}

The tip-position dependence of the voltage drop is modeled using a two-point-charge geometry as depicted in figure~\ref{fig:simMaps_VD}c.
The electric potential in the junction, induced by the applied bias voltage $V_b$ is approximated as $\phi_b(\mathbf{r}) = \varphi (\mathbf{r}) V_b$, where
\begin{equation}
    \varphi(\mathbf{r}) = 1- \left ( \frac{R^2}{2z_t} + R \right )\left[ 
    \frac{1}{\sqrt{x^2+y^2+(z-z_p)^2}} -
    \frac{1}{\sqrt{x^2+y^2+(z+z_p)^2}}
    \right],
\end{equation}
with $R=z_p-z_t$ denoting the effective tip radius~\cite{Krane2019}. The effective potential applied to the molecule is then given by
\begin{equation}
    V_m(\mathbf{r_t}) = V_b \int \mathrm{d}\mathbf{r} \Psi^\ast(\mathbf{r}) \varphi(\mathbf{r}-\mathbf{r_t}) \Psi(\mathbf{r}),
\end{equation}
with $\Psi$ representing the wave function of the corresponding Dyson orbital. In good approximation, we use the single-particle wave functions of \TBP\ as $\Psi$. 

Finally, the effective voltage drop $\alpha$ is defined as the ratio between the applied bias $V_b$ and the effective potential $V_m$ applied to the molecule:
\begin{equation}
    \alpha(\mathbf{r}) = 1 - \frac{V_m}{V_b} = 1 - \int \mathrm{d}\mathbf{r} \Psi^\ast(\mathbf{r}) \varphi(\mathbf{r}-\mathbf{r_t}) \Psi(\mathbf{r}).
\end{equation}
The spatial dependence of the voltage drop for the HOMO and the HOMO-1 is shown in Figure~\ref{fig:simMaps_VD}d,e for two point charges, located at $z_p=\pm20$\,\AA, and a vacuum gap of $z_t=10$\,Å. The molecule is displaced by $z_m=1$\,\AA\ from the surface plane to account for the dielectric decoupling layer. The junction geometry was chosen such that the voltage drop reaches $\alpha\approx10$\,\% at the position of maximum tip--molecule coupling for the HOMO orbital. Figure~\ref{fig:simMaps_VD}f shows two line cuts of the simulated voltage drop, illustrating that the voltage drop is largest when the tip is directly above the molecule and decreases with increasing lateral distance.

\begin{figure}[htbp]
    \centering
    \includegraphics[width=0.6\linewidth]{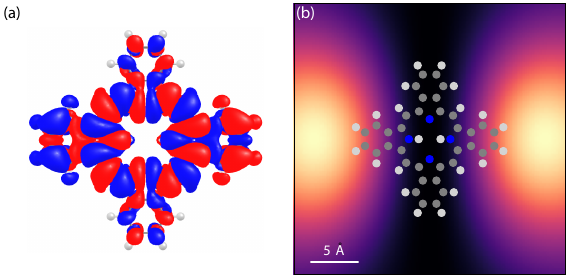}
    \caption{Transition density of the \ESa\ state of \TBP. (a) iso-surface of the transition density mapped onto stick-and-ball model of \TBP. (b) Radiative coupling strength $\hat{V}_\mathrm{rad}(\mathbf{r_t})$ according to eq.~\eqref{eq:transdens}.}
    \label{fig:simMaps_TD}
\end{figure}

The tip plasmon is modeled using the same two-point-charge geometry described above. The coupling strength to the transition density $\rho_{if}$ between states $i$ to $f$ is then given by~\cite{Ferreira2025}:
\begin{equation}
    \hat{V}_\mathrm{rad}(\mathbf{r_t}) = \hat{V}_\mathrm{rad}^0 \frac{\left | g(\mathbf{r_t})\right|^2 }{
    \mathrm{max} \left ( \left | g(\mathbf{r_t})\right|^2 \right )},
    \label{eq:transdens}
\end{equation}
with $\hat{V}_\mathrm{rad}^0=1$\,meV and
\begin{equation}
    g(\mathbf{r_t}) = \int \mathrm{d}\mathbf{r} \rho_{ij}(\mathbf{r}) \phi(\mathbf{r}-\mathbf{r_t}).
\end{equation}
The tip potential $\phi(\mathbf{r})$ is described by the electrostatic potential of the two point charges as depicted in Figure~\ref{fig:simMaps_VD}e:
\begin{equation}
    \phi(\mathbf{r}) =  \frac{1}{\sqrt{x^2+y^2+(z-z_p)^2}} -
    \frac{1}{\sqrt{x^2+y^2+(z+z_p)^2}}.
\end{equation}
Figure~\ref{fig:simMaps_TD} shows the transition density $\rho_{if}$ for the \ESa\ state of \TBP, together with the corresponding spatial map of the plasmon coupling strength $\hat{V}_\mathrm{rad}(\mathbf{r_t})$.

\begin{figure}[htbp]
    \centering
    \includegraphics[width=0.6\linewidth]{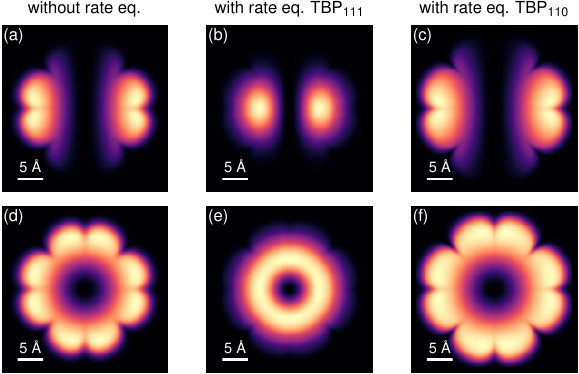}
    \caption{Simulated flourescence maps for \TBP.
    (a) Map obtained using the simple model (Eq.~\ref{eq:simple_map}), without the rate equation model or the effects of voltage drop.
    (b,c) Fluorescence maps calculated using the rate-equation model, including a spatially dependent voltage drop, for \To\ and \Tz, respectively.
    (d-f) Fluorescence maps including tautomerization effects, constructed by rotating each map in (a–c) by 90° and summing the resulting pair.
    }
    \label{fig:simMaps_maps}
\end{figure}

For systems that are not affected by spatially dependent gating, normalized STML maps can be reproduced by multiplying the tip-plasmon coupling strength $\left|g(\mathbf{r_t})\right|^2$ with the tip coupling rate $\hat{V}_t(\mathbf{r_t})$ and subsequent renormalization, while including an additional background current~\cite{roslawska2022mapping}:
\begin{equation}
    j_\mathrm{ph}^\mathrm{sim}(\mathbf{r}) \propto \left | g(\mathbf{r_t})\right|^2 \frac{\hat{V}_t(\mathbf{r_t})}{
    \hat{V}_t(\mathbf{r_t}) + \gamma_\mathrm{bg}},
    \label{eq:simple_map}
\end{equation}
with $\gamma_\mathrm{bg} = 0.01\hat{V}_t^0$. The resulting map is shown in Figure~\ref{fig:simMaps_maps}a for the \ESa\ state of \TBP. Figure~\ref{fig:simMaps_maps}d shows the corresponding map, when fast tautomerization is assumed, with an equal weighting of the two tautomers.

To capture the effect of spatially dependent voltage drop, we simulate the STML maps by solving the rate equation at each tip position, using the definitions for $\hat{V}_t(\mathbf{r})$, $\hat{V}_\mathrm{rad}(\mathbf{r})$, and  $\alpha(\mathbf{r})$ introduced above.
 The resulting fluorescence map is then normalized by the current, including a dark background current of $I_\mathrm{bg}=150$\,fA:
\begin{equation}
    j_\mathrm{ph}^\mathrm{RE}(\mathbf{r}) = \frac{J_\mathrm{ph}(\mathbf{r})}{I_\mathrm{mol}(\mathbf{r}) + I_\mathrm{bg}}.
    \label{eq:re_map}
\end{equation}
Figure~\ref{fig:simMaps_maps}b,e and \ref{fig:simMaps_maps}c,f show the simulated STML maps for \To\ and \Tz, respectively, both without and with the inclusion of tautomerization. The simulated maps for \Tz\ are nearly identical to those obtained from the simplified model shown in Figure~\ref{fig:simMaps_maps}a,d, supporting the conclusion that \Tz\ is largely insensitive to the gating effect. In contrast, \To\ deviates from the idealized model, exhibiting a ring-shaped intensity distribution with the maximum shifted towards the center of the molecule. This result is consistent with the experimental maps (Figure 4) and highlights that the inclusion of the gating effect introduces an additional layer of complexity in interpretation of spatially resolved STML data.

\clearpage
\bibliography{references}